\documentclass[aps,prstab,superscriptaddress,longbibliography, twocolumn,amsmath,amssymb]{revtex4-2}	
\usepackage{graphicx}
\usepackage{dcolumn} 
\usepackage{hyperref}
\hypersetup{
    colorlinks=true,        
    linkcolor=blue,          
    citecolor=blue,        
    filecolor=magenta,      
    urlcolor=blue           
}
\newcommand{\Figref}[1]{Fig.\ref{#1}}

\newcommand{\citenamefont}{}
\newcommand{\bibnamefont}{}

\usepackage{color} 				
\usepackage[normalem]{ulem} 			

\begin{document}
\title{Theory of wakefields excited by an off--axis drive bunch in a plasma--dielectric waveguide}

\author{K.V. Galaydych}\email{kgalaydych@gmail.com}
\affiliation{NSC Kharkiv Institute of Physics and Technology, 61108 Kharkiv, Ukraine}
\author{G.V. Sotnikov}
\affiliation{NSC Kharkiv Institute of Physics and Technology, 61108 Kharkiv, Ukraine}
\author{I.N. Onishchenko}
\affiliation{NSC Kharkiv Institute of Physics and Technology, 61108 Kharkiv, Ukraine}

\date{\today}

\begin{abstract}
A linear theory of wakefield excitation by a drive electron bunch in a plasma--dielectric accelerating structure has been constructed for the case of off--axis bunch injection. The structure under investigation is a round dielectric--loaded metal waveguide with a channel for charged particles, filled with homogeneous cold plasma. The developed theory was used to investigate the spatial distribution of the bunch--excited transverse wakefield components, which act on the drive electron bunch. The numerical simulations using the particle-in-cell method have confirmed the predictions of the linear theory in terms of transverse drive bunch dynamics. The numerical analysis has shown stable drive bunch transport through the plasma--dielectric accelerating structure compared with the corresponding dielectric--loaded structure.
\end{abstract}
\maketitle

\section{INTRODUCTION}
Among the advanced acceleration concepts it is possible to highlight four main schemes, namely, beam--driven plasma wakefield acceleration~\cite{Hogan2016,Adli2016}, laser--driven plasma wakefield acceleration~\cite{Nakajima2016}, dielectric laser acceleration~\cite{Wootton2016} and structure--based wakefield acceleration~\cite{Jing2016}. The dielectric wakefield accelerating structures belong to the last one and for the present time a significant progress in their research and development has been made (see, for example, papers~\cite{Thompson2008,O'Shea2016,Gao2018}). These structures are in focus of intense studies at such accelerator centers as ANL, SLAC, BNL, INFN. But despite the efforts and progress, there are the important tasks and problems that still should be solved. And solving them may require nonconventional methods and approaches. As for the conventional accelerators, one of the severe issues of the advanced accelerators development is the beam breakup (BBU) instability of the drive bunch. The studies of this instability are of vital importance because this process is destructive and causes degradation of the drive bunch and the generated wakefield quality, reduces the drive bunch intensity, and restricts its propagation length. In turn, all these negative factors can affect the charged particle acceleration parameters provided by the dielectric wakefield accelerating structures. The beam breakup problem has been known and studied for a long time both by theory~\cite{Loew1967,Panofsky1968,Neil1979} and by experiment~\cite{Altenmueller1966,Caporaso1983}, and has been described in numerous publications. In~\cite{Lau1989} Lau has presented a classification of beam breakup instabilities in linacs. The previous studies~\cite{Ng,RosingGai,Garate,GaiKanareykin,ParkHirshfield2000,ParkHirshfield2001,Li,Lebedev}, devoted to the transverse wakefields, and, in particular, to the BBU directly in the dielectric wakefield accelerating structures, point out a great importance of this problem, which is still not completely overcome at the moment. To control and suppress this instability, several methods have been proposed. Among them are (I) the Balakin--Novokhatsky--Smirnov (BNS) damping~\cite{Balakin}, which consists in variation of the betatron oscillation frequency along the bunch; (II) the combination of a profiled quadrupole and FODO focusing system with an energy chirp of bunch ~\cite{Shchegolkov2016,BaturinZholents}; (III) the excitation of a dielectric resonator by a train of bunches~\cite{Galaydych}. In~\cite{Tremaine1997,BaturinAndonian2018,OShea2020} a transverse wakefield reduction using the elliptical (transversally) drive bunches was investigated theoretically and experimentally.
On the other hand in papers~\cite{Ruth,Rosenzweig} it has been demonstrated that plasma has the focusing properties, and as a result, this allows one to focus the drive bunch. The authors of paper~\cite{Sotnikov2014} have demonstrated that a combination of a dielectric--loaded structure and plasma provides intrinsic focusing features and causes the test and drive bunches to be focused. It has been shown that in this plasma--dielectric structure, the acceleration is provided by the $TM$ eigenwave of the structure, and the focusing is provided by the plasma wave. Moreover, further research~\cite{MarkovJINST,sotnikov:ipac2021-mopab145} has shown that using a nonuniform radial plasma density profile, which takes place at capillary discharge, improves focusing of both the drive and witness bunches, and also there exists the optimum vacuum channel radius, at which the focusing is the best.
 So, the mentioned features of the plasma--dielectric structure allow one to consider it as an alternative to a standard quadrupole focusing system, and as a possible way for the beam breakup mitigation during wakefield excitation and acceleration. This was the primary motivation of our studies and the present paper. Here we have centred on the analytical studies of wakefields generation by the off--axis drive bunch in the plasma--dielectric waveguide. We have focused our attention on the closed--form expressions for the bunch--excited electric and magnetic field components in the rigid drive bunch approximation. The paper is organized as follows. Sec. II gives the statement of the problem, Sec. III is devoted to the analytical bunch--excited wakefield studies, Sec. IV demonstrates the results of the numerical studies and Sec. V gives a short summary.

\section{STATEMENT OF THE PROBLEM}
\begin{figure}[!th]
  \centering
  \includegraphics[width=0.5\textwidth]{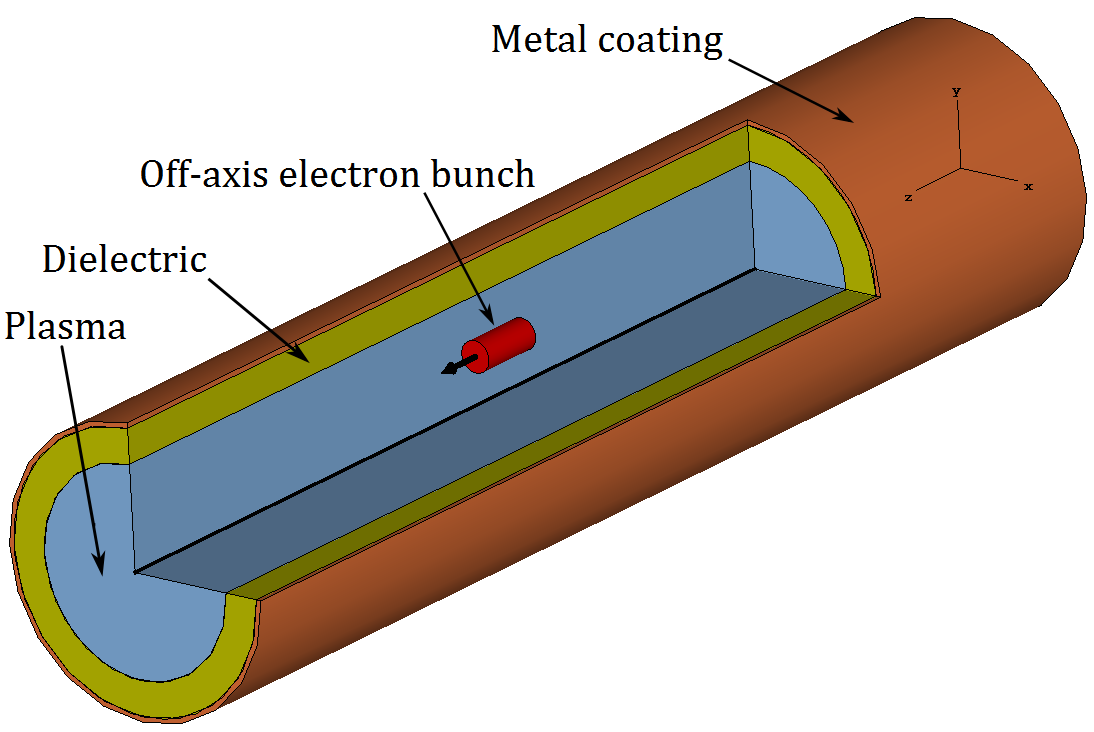}
  \caption{General view of the round plasma--dielectric waveguide. Metal coating (orange), dielectric (yellow), plasma (blue) and electron bunch (red) are shown schematically. The off--axis electron bunch moves in parallel to the axis.}\label{Fig:01}
\end{figure}

The accelerating structure is a round dielectric--loaded metal waveguide with a channel for charged particles. This channel is filled with homogeneous cold plasma. A relativistic drive electron bunch is injected with offset into plasma in parallel to the structure axis. The wakefield is excited by the drive bunch passing through the structure. For the case of on--axis injection, the excited wakefield is described in terms of TM eigenmodes only. However, due to the drive bunch offset, the excited wakefield is not described only in terms of TM eigenmodes and has all six components of the electromagnetic field. A three--dimensional general schematic of the plasma--dielectric accelerating structure with the drive bunch passing through is shown in \Figref{Fig:01}.
The main goal of the paper is to construct a linear theory of wakefield excitation in the plasma--dielectric waveguide by the drive electron bunch for the case of off--axis bunch injection taking into account the high--order azimuthal modes.
\section{ANALYTICAL STUDIES}
The wakefield excited by the point--like charged particle (Green's function) should be analyzed first. For the charged particle $q$ moving with the constant velocity $v$ along the waveguide axis ($z$ direction) with the transverse coordinates $r_0$ and $\varphi_0$, the charge and current densities in the cylindrical coordinates are as follows:
\begin{equation}\label{eq:01}
\begin{split}
{\rho} = \frac{q}{v}\frac{{\delta (r - {r_0})}}{r}\delta (\varphi  - {\varphi _0})\delta (t - {t_0} - z/v),\\
{j_z} = q\frac{{\delta (r - {r_0})}}{r}\delta (\varphi  - {\varphi _0})\delta (t - {t_0} - z/v),
\end{split}
\end{equation}
where $t_0$ is the time of particle arrival into the waveguide ($z=0$), and $\delta$ is the Dirac delta function. As the longitudinal boundedness effect is neglected, it is supposed that the excited wakefield is defined by the source only. Therefore, all the electromagnetic field components depend on the axial coordinate $z$ and time $t$ similar to the functions of $t-z/v$. In this paper, the transient process of electromagnetic field formation and group velocity effects associated with the longitudinal boundedness of the practicable accelerating structures are not taken into account. The bunch--excited wakefield components, as well as the charge and current densities, can be described in terms of Fourier transforms by the variables $t - z/v$ and $\varphi$ as follows:
\begin{equation}\label{eq:02}
\begin{split}
{\bf{E}}(r,\varphi ,\xi) &= \sum\limits_{m =  - \infty }^{ + \infty } {{e^{im\varphi }}\int\limits_{ - \infty }^{ + \infty } {d\omega {\bf{E}}_m^\omega (r,\omega )} } {e^{ - i\omega \xi}}{\text{,}}\\
{\bf{D}}(r,\varphi ,\xi) &= \sum\limits_{m =  - \infty }^{ + \infty } {{e^{im\varphi }}\int\limits_{ - \infty }^{ + \infty } {d\omega \varepsilon (\omega ){\bf{E}}_m^\omega (r,\omega )} } {e^{ - i\omega \xi}}{\text{,}}\\
{\bf{H}}(r,\varphi ,\xi) &= \sum\limits_{m =  - \infty }^{ + \infty } {{e^{im\varphi }}\int\limits_{ - \infty }^{ + \infty } {d\omega {\bf{H}}_m^\omega (r,\omega )} } {e^{ - i\omega \xi}}{\text{,}}\\
{\bf{B}}(r,\varphi ,\xi) &= \sum\limits_{m =  - \infty }^{ + \infty } {{e^{im\varphi }}\int\limits_{ - \infty }^{ + \infty } {d\omega \mu (\omega ){\bf{H}}_m^\omega (r,\omega )} } {e^{ - i\omega \xi}}{\text{,}}\\
{\rho}(r,\varphi ,\xi) &= \sum\limits_{m =  - \infty }^{ + \infty } {{e^{im\varphi }}\int\limits_{ - \infty }^{ + \infty } {d\omega {\rho}_{m}^\omega (r,\omega )} } {e^{ - i\omega \xi}}{\text{,}}\\
{j_z}(r,\varphi ,\xi) &= \sum\limits_{m =  - \infty }^{ + \infty } {{e^{im\varphi }}\int\limits_{ - \infty }^{ + \infty } {d\omega {j}_{zm}^\omega (r,\omega )} } {e^{ - i\omega \xi}}{\text{,}}
\end{split}
\end{equation}
where $\xi = t - z/v$. The set of Maxwell's equations results in the system of coupled equations for the Fourier transforms of all the electric and magnetic field components $E_{rm}^\omega$, $E_{\varphi m}^\omega$, $E_{zm}^\omega$, $H_{rm}^\omega$, $H_{\varphi m}^\omega$ and $H_{zm}^\omega$. In turn, the aforementioned system of the coupled equations leads to the uncoupled wave equations for the Fourier transforms of the longitudinal electric and magnetic fields $E_{zm}^\omega$, and $H_{zm}^\omega$, namely:
\begin{equation}\label{eq:03}
\begin{split}
&\frac{1}{r}\frac{\partial }{{\partial r}}\left( {r\frac{{\partial E_{zm}^\omega }}{{\partial r}}} \right) - \frac{{{m^2}}}{{{r^2}}}E_{zm}^\omega  - \frac{{{\omega ^2}}}{{{v^2}}}(1 - {\beta ^2}\varepsilon (\omega ))E_{zm}^\omega  = \\
&\frac{{4\pi \omega i}}{{{v^2}}}\,\frac{{1 - {\beta ^2}\varepsilon (\omega )}}{{\varepsilon (\omega )}}j_{zm}^\omega ,\\
&\frac{1}{r}\frac{\partial }{{\partial r}}\left( {r\frac{{\partial H_{zm}^\omega }}{{\partial r}}} \right) - \frac{{{m^2}}}{{{r^2}}}H_{zm}^\omega  - \frac{{{\omega ^2}}}{{{v^2}}}(1 - {\beta ^2}\varepsilon (\omega ))H_{zm}^\omega  = 0.
\end{split}
\end{equation}

All the other four Fourier transforms $E_{rm}^\omega$, $E_{\varphi m}^\omega$, $H_{rm}^\omega$ and $H_{\varphi m}^\omega$ can be described in terms of $E_{zm}^\omega$, and $H_{zm}^\omega$.

\begin{equation}\label{eq:04}
\begin{split}
E_{rm}^\omega  &= \frac{v/\omega }{{(1 - {\beta ^2}\varepsilon (\omega ))}}\left( {\beta \frac{m}{r}H_{zm}^\omega  - i\frac{{\partial E_{zm}^\omega }}{{\partial r}}} \right),\\
E_{\varphi m}^\omega  &= \frac{{v/\omega }}{{(1 - {\beta ^2}\varepsilon (\omega ))}}\left( {\frac{m}{r}E_{zm}^\omega  + i\beta\frac{{\partial H_{zm}^\omega }}{{\partial r}}} \right),\\
H_{rm}^\omega  &= \frac{v/\omega }{{(1 - {\beta ^2}\varepsilon (\omega ))}}\left( { - i\frac{{\partial H_{zm}^\omega }}{{\partial r}} - \beta \frac{m}{r}\varepsilon (\omega )E_{zm}^\omega } \right),\\
H_{\varphi m}^\omega  &= \frac{v/\omega }{{(1 - {\beta ^2}\varepsilon (\omega ))}}\left( {\frac{m}{r}H_{zm}^\omega  - i\beta \varepsilon (\omega )\frac{{\partial E_{zm}^\omega }}{{\partial r}}} \right).
\end{split}
\end{equation}

To get the expressions for the Fourier transforms, the boundary conditions for the waveguide metal coating and plasma--dielectric interface should be added. These boundary conditions are as follows: (I) for the waveguide wall ($r=b$), the axial components of the electric field ($E_{zm}^{\omega}$) should be equal to zero; (II) for the plasma--dielectric interface ($r=a$), the axial components of the electric field ($E_{zm}^{\omega}$, $E_{\varphi m}^{\omega}$), axial components of the magnetic induction ($B_{zm}^{\omega}$) and radial components of the electric induction ($D_{rm}^{\omega}$) should be continuous. The boundary conditions for the rest of the wakefield components will be satisfied automatically. In this paper, mathematical details will be omitted and the results for the axial components of electric and magnetic fields will be presented for any given azimuthal $m^{th}$ mode in the plasma and dielectric regions. A detailed description of the procedure for constructing the solutions for $E_{zm}^{\omega}$ and $H_{zm}^{\omega}$ can be found, for example, in papers ~\cite{Ng,RosingGai,Garate,ParkHirshfield2000}. The results satisfied by the aforementioned boundary condition are as follows:

\begin{equation}\label{eq:05}
\begin{split}
&E_{zm}^\omega (r < {r_0}) =  - \frac{{iq}}{\pi }\left( {\frac{{{D_2}(\omega )}}{{a\omega D(\omega )}}} \right.\frac{{{I_m}({\kappa _p}r){I_m}({\kappa _p}{r_0})}}{{I_m^2({\kappa _p}a)}} +\\ &\left. {\frac{{\kappa _p^2}}{{\omega {\varepsilon _p}(\omega )}}\frac{{{I_m}({\kappa _p}r)}}{{{I_m}({\kappa _p}a)}}{\Delta _m}({\kappa _p}a,{\kappa _p}{r_0})} \right){e^{ - im{\varphi _0} + i\omega {t_0}}},\\
&E_{zm}^\omega (r > {r_0}) =  - \frac{{iq}}{\pi }\left( {\frac{{{D_2}(\omega )}}{{a\omega D(\omega )}}} \right.\frac{{{I_m}({\kappa _p}r){I_m}({\kappa _p}{r_0})}}{{I_m^2({\kappa _p}a)}} +\\ &\left. {\frac{{\kappa _p^2}}{{\omega {\varepsilon _p}(\omega )}}\frac{{{I_m}({\kappa _p}{r_0})}}{{{I_m}({\kappa _p}a)}}{\Delta _m}({\kappa _p}a,{\kappa _p}r)} \right){e^{ - im{\varphi _0} + i\omega {t_0}}},\\
&E_{zm}^\omega (a < r < b) =  - \frac{{iq}}{\pi }\frac{{{D_2}(\omega )}}{{a\omega D(\omega )}}\frac{{{I_m}({\kappa _p}{r_0})}}{{{I_m}({\kappa _p}a)}}\times\\
&\frac{{{F_m}({\kappa _d}r,{\kappa _d}b)}}{{{F_m}({\kappa _d}a,{\kappa _d}b)}}{e^{ - im{\varphi _0} + i\omega {t_0}}},
\end{split}
\end{equation}

\begin{equation}\label{eq:06}
\begin{split}
&H_{zm}^\omega (0 < r \le a) = \frac{q}{\pi }\frac{\beta m}{{a^2\omega D(\omega )}}\frac{{{{(\omega /v)}^2}({\varepsilon _d} - {\varepsilon _p}(\omega ))}}{{\kappa _p^2\kappa _d^2}}\times\\
&\frac{{{I_m}({\kappa _p}r){I_m}({\kappa _p}{r_0})}}{{I_m^2({\kappa _p}a)}}{e^{ - im{\varphi _0} + i\omega {t_0}}},\\
&H_{zm}^\omega (a < r \le b) = \frac{q}{\pi }\frac{\beta m}{{a^2\omega D(\omega )}}\frac{{{{(\omega /v)}^2}({\varepsilon _d} - {\varepsilon _p}(\omega ))}}{{\kappa _p^2\kappa _d^2}}\times\\
&\frac{{{I_m}({\kappa _p}{r_0})}}{{{I_m}({\kappa _p}a)}}\frac{{{\Phi _m}({\kappa _d}r,{\kappa _d}b)}}{{{\Phi _m}({\kappa _d}a,{\kappa _d}b)}}{e^{ - im{\varphi _0} + i\omega {t_0}}},
\end{split}
\end{equation}
where the following notations are used: $a$ is the inner radius of the dielectric, $\kappa _{p}^2 = (\omega/v)^2(1 - {\beta ^2}{\varepsilon _p}({\omega}))$, $\kappa_d^2=(\omega/v)^2(\beta^2\varepsilon_d-1)$, ${\Delta _m}(x,y) = {I_m}(x){K_m}(y) - {K_m}(x){I_m}(y)$, $I_m$ and $K_m$ are the modified Bessel and Macdonald functions of the $m^{th}$ order, ${\varepsilon _p}(\omega ) = 1 - \omega _p^2/{\omega ^2}$ is the plasma permittivity, $\omega_p$ is the plasma frequency, $\varepsilon_d$ is the dielectric permittivity, ${F_m}(x,y) = {( - 1)^m}({J_m}(x){Y_m}(y) - {Y_m}(x){J_m}(y))$, $J_m$ and $Y_m$ are the Bessel and Weber functions of the order $m^{th}$, ${F'_m}(x,y) = {( - 1)^m}({J'_m}(x){Y_m}(y) - {Y'_m}(x){J_m}(y))$, ${\Phi _m}(x,y) = {J_m}(x){Y'_m}(y) - {Y_m}(x){J'_m}(y)$, ${\Phi '_m}(x,y) = {J'_m}(x){Y'_m}(y) - {Y'_m}(x){J'_m}(y)$. In the expressions ~(\ref{eq:05}) and ~(\ref{eq:06}) the function $D(\omega)$ is a dispersion relation and it has the following form

\begin{equation}\label{eq:07}
\begin{split}
D(\omega) &= D_1(\omega)D_2(\omega) - \frac{\beta^2m^2(\omega/v)^4}{a^2\kappa_p^4\kappa_d^4}(\varepsilon_d-\varepsilon_p)^2,\\
D_1(\omega) &= \frac{\varepsilon_p}{\kappa_p}\frac{I^\prime_m(\kappa_p a)}{I_m(\kappa_p a)} + \frac{\varepsilon_d}{\kappa_d}\frac{F^\prime_m(\kappa_d a, \kappa_d b)}{F_m(\kappa_d a, \kappa_d b)},\\
D_2(\omega) &= \frac{1}{\kappa_p}\frac{I^\prime_m(\kappa_p a)}{I_m(\kappa_p a)} + \frac{1}{\kappa_d}\frac{\Phi^\prime_m(\kappa_d a, \kappa_d b)}{\Phi_m(\kappa_d a, \kappa_d b)}.
\end{split}
\end{equation}

The frequencies of the plasma--dielectric waveguide eigenmodes, which are in Cherenkov resonance with the drive bunch, are defined by the roots of the equation $D(\omega)=0$. For the case of azimuthaly homogenous eigenmodes ($m=0$), these frequencies are defined from the dispersion equation of the resonant $TM$--modes as follows:

\begin{equation}\label{eq:08}
\begin{split}
\frac{{{\varepsilon _p}(\omega )}}{{{\kappa _p}}}\frac{{{{I'}_0}({\kappa _p}a)}}{{{I_0}({\kappa _p}a)}} + \frac{{{\varepsilon _d}}}{{{\kappa _d}}}\frac{{{{F'}_0}({\kappa _d}a,{\kappa _d}b)}}{{{F_0}({\kappa _d}a,{\kappa _d}b)}} = 0.
\end{split}
\end{equation}

Due to the inverse Fourier transform of the expressions ~(\ref{eq:05}) and ~(\ref{eq:06}), the space--time distribution of the wakefield components excited by the point--like charged particle in the plasma--dielectric waveguide can be obtained. It should be noted that the expressions ~(\ref{eq:05}) and ~(\ref{eq:06}) have simple poles $\varepsilon_p(\omega)=0$ and $D(\omega)=0$ only. The wakefield of plasma wave is defined by the poles $\varepsilon_p(\omega)=0$. The wakefield of the plasma--dielectric eigenmodes, which are synchronous with the drive bunch, is defined by the poles $D(\omega)=0$, as was mentioned above. The inverse Fourier transforms with respect to the frequency, using of the calculus of residues, gives for $E_{zm}$ and $H_{zm}$, the following:

\begin{equation}\label{eq:9}
\begin{split}
&{E_{zm}}(r < {r_0},\varphi ,\xi ,{r_0},{\varphi _0},{t_0}) =  - 2qk_p^2\frac{{{I_m}({k_p}r)}}{{{I_m}({k_p}a)}}\times\\
&{\Delta _m}({k_p}a,{k_p}{r_0})\cos {\omega _p}(\xi  - {t_0})\theta (\xi  - {t_0}){e^{ - im{\varphi _0}}} - \\
&\sum\limits_{s = 1} {\frac{{4q{D_2}({\omega _s})}}{{a{\omega _s}D'({\omega _s})}}} \frac{{{I_m}({\kappa _{ps}}r){I_m}({\kappa _{ps}}{r_0})}}{{I_m^2({\kappa _{ps}}a)}}\times\\
&\cos {\omega _s}(\xi  - {t_0})\theta (\xi  - {t_0}){e^{ - im{\varphi _0}}},\\
\end{split}
\end{equation}
\begin{equation}\label{eq:10}
\begin{split}
&{E_{zm}}({r_0} < r < a,\varphi ,\xi ,{r_0},{\varphi _0},{t_0}) =  - 2qk_p^2\frac{{{I_m}({k_p}{r_0})}}{{{I_m}({k_p}a)}}\times\\
&{\Delta _m}({k_p}a,{k_p}r)\cos {\omega _p}(\xi  - {t_0})\theta (\xi  - {t_0}){e^{ - im{\varphi _0}}} -\\
&\sum\limits_{s = 1} {\frac{{4q{D_2}({\omega _s})}}{{a{\omega _s}D'({\omega _s})}}} \frac{{{I_m}({\kappa _{ps}}r){I_m}({\kappa _{ps}}{r_0})}}{{I_m^2({\kappa _{ps}}a)}}\times\\
&\cos {\omega _s}(\xi  - {t_0})\theta (\xi  - {t_0}){e^{ - im{\varphi _0}}},
\end{split}
\end{equation}
\begin{equation}\label{eq:11}
\begin{split}
&{E_{zm}}(a < r \le b,\varphi ,\xi ,{r_0},{\varphi _0},{t_0}) =  - \sum\limits_{s = 1} {\frac{{4q{D_2}({\omega _s})}}{{a{\omega _s}D'({\omega _s})}}}\times\\
&\frac{{{I_m}({\kappa _{ps}}{r_0})}}{{{I_m}({\kappa _{ps}}a)}}\frac{{{F_m}({\kappa _{ds}}r,{\kappa _{ds}}b)}}{{{F_m}({\kappa _{ds}}a,{\kappa _{ds}}b)}}\cos {\omega _s}(\xi  - {t_0})\theta (\xi  - {t_0}){e^{ - im{\varphi _0}}},
\end{split}
\end{equation}
\begin{equation}\label{eq:12}
\begin{split}
&H_{zm}(0 < r \le a,\varphi ,\xi ,{r_0},{\varphi _0},{t_0}) =  - \sum\limits_{s = 1} {i\frac{{4q\beta m}}{{{a^2}{\omega _s}D'({\omega _s})}}}  \times \\
&\frac{{{I_m}({\kappa _{ps}}r){I_m}({\kappa _{ps}}{r_0})}}{{I_m^2({\kappa _{ps}}a)}}\frac{{{{({\omega _s}/v)}^2}}}{{\kappa _{ps}^2\kappa _{ds}^2}}({\varepsilon _d} - {\varepsilon _p}({\omega _s})) \times \\
&\cos {\omega _s}(\xi  - {t_0})\theta (\xi  - {t_0}){e^{ - im{\varphi _0}}},
\end{split}
\end{equation}
\begin{equation}\label{eq:13}
\begin{split}
&H_{zm}(a < r \le b,\varphi ,\xi ,{r_0},{\varphi _0},{t_0}) =  - \sum\limits_{s = 1} {i\frac{{4q\beta m}}{{{a^2}{\omega _s}D'({\omega _s})}}}  \times \\
&\frac{{{I_m}({\kappa _{ps}}r)}}{{{I_m}({\kappa _{ps}}a)}}\frac{{{\Phi _m}({\kappa _{ds}}r,{\kappa _{ds}}b)}}{{{\Phi _m}({\kappa _{ds}}a,{\kappa _{ds}}b)}}\frac{{{{({\omega _s}/v)}^2}}}{{\kappa _{ps}^2\kappa _{ds}^2}}({\varepsilon _d} - {\varepsilon _p}({\omega _s})) \times \\
&\cos {\omega _s}(\xi  - {t_0})\theta (\xi  - {t_0}){e^{ - im{\varphi _0}}},
\end{split}
\end{equation}
In this paper, the structure of the longitudinal and radial forces acting on the bunch particles are of vital interest. Therefore, the expressions for the axial and radial components of the electric field, as well as for the azimuthal component of the magnetic field only, are presented as follows:
\begin{equation}\label{eq:14}
\begin{split}
&{E_{rm}}(r < {r_0},\varphi ,\xi ,{r_0},{\varphi _0},{t_0}) = 2qk_p^2\frac{{I'_m({k_p}r)}}{{{I_m}({k_p}a)}}\times\\
&\Delta _m({k_p}a,{k_p}{r_0})\sin {\omega _p}(\xi  - {t_0})\theta (\xi  - {t_0}){e^{ - im{\varphi _0}}}-\\
&\sum\limits_{s = 1} {\frac{{4q}}{{a{\kappa _{ps}}vD'({\omega _s})}}} \frac{{{I_m}({\kappa _{ps}}{r_0})}}{{I_m^2({\kappa _{ps}}a)}}\times\\
&\left({ \frac{{{m^2}{\beta ^2}k_s^2}}{{a\kappa _{ps}^2\kappa _{ds}^2}}({\varepsilon _d} - {\varepsilon _p}({\omega _s}))\frac{{{I_m}({\kappa _{ps}}r)}}{{{\kappa _{ps}}r}} - {D_2}({\omega _s})I'_m({\kappa _{ps}}r)}\right)\times\\
&\sin {\omega _s}(\xi  - {t_0})\theta (\xi  - {t_0}){e^{ - im{\varphi _0}}},
\end{split}
\end{equation}

\begin{equation}\label{eq:15}
\begin{split}
&{E_{rm}}({r_0} < r < a,\varphi ,\xi ,{r_0},{\varphi _0},{t_0}) = 2qk_p^2\frac{{{I_m}({k_p}{r_0})}}{{{I_m}({k_p}a)}}\times\\
&\Delta '_m({k_p}a,{k_p}r)\sin {\omega _p}(\xi  - {t_0})\theta (\xi  - {t_0}){e^{ - im{\varphi _0}}}-\\
&\sum\limits_{s = 1} {\frac{{4q}}{{a{\kappa _{ps}}vD'({\omega _s})}}} \frac{{{I_m}({\kappa _{ps}}{r_0})}}{{I_m^2({\kappa _{ps}}a)}}\times\\
&\left({\frac{{{m^2}{\beta ^2}k_s^2}}{{a\kappa _{ps}^2\kappa _{ds}^2}}({\varepsilon _d} - {\varepsilon _p}({\omega _s}))\frac{{{I_m}({\kappa _{ps}}r)}}{{{\kappa _{ps}}r}} - {D_2}({\omega _s})I'_m({\kappa _{ps}}r)}\right)\times\\
&\sin {\omega _s}(\xi  - {t_0})\theta (\xi  - {t_0}){e^{ - im{\varphi _0}}},
\end{split}
\end{equation}

\begin{equation}\label{eq:16}
\begin{split}
&{H_{\varphi m}}(0 < r \le a,\varphi ,\xi ,{r_0},{\varphi _0},{t_0}) =  - \sum\limits_{s = 1} {\frac{{4q}}{{a{\kappa _{ps}}cD'({\omega _s})}}}  \times \\
&\frac{{{I_m}({\kappa _{ps}}{r_0})}}{{I_m^2({\kappa _{ps}}a)}} \bigg({\frac{{{m^2}k_s^2}}{{a\kappa _{ps}^2\kappa _{ds}^2}}({\varepsilon _d} - {\varepsilon _p}({\omega _s}))\frac{{{I_m}({\kappa _{ps}}r)}}{{{\kappa _{ps}}r}} - } \\
& {{\varepsilon _p}({\omega _s}){D_2}({\omega _s})I'_m({\kappa _{ps}}r)} \bigg)\sin {\omega _s}(\xi  - {t_0})\theta (\xi  - {t_0}){e^{ - im{\varphi _0}}}
\end{split}
\end{equation}

where ${\kappa _{ps}} = {\kappa _p}({\omega _s})$, ${\kappa _{ds}} = {\kappa _d}({\omega _s})$, $D'({\omega _s})=dD({\omega_s})/d\omega_s$, and  $\theta$ is the Heaviside function. The space--time structure of the wakefield excited by a point--like charged particle injected with offset in the waveguide under study is described by the expressions ~(\ref{eq:9})--~(\ref{eq:16}). As it can be seen from the expressions ~(\ref{eq:9}),~(\ref{eq:10}) and ~(\ref{eq:14}),~(\ref{eq:15}) for the longitudinal and radial components of the electric field in the plasma region, the field of plasma wave is described by the first terms. This field is potential and localized in the plasma--filled channel for charged particles. The longitudinal and radial electric fields of the eigenwaves of the plasma--dielectric waveguide, are described by the other terms, respectively. The azimuthal component of the magnetic field does not include the plasma wave wakefield. In the process of wakefield excitation, the longitudinal dynamics of the bunch is determined by the axial component of the electric field ($F_z/q=E_z$), and the radial dynamics is determined by the radial component of the Lorentz force, which is shown as $F_r/q=E_r-\beta B_\varphi$. In the plasma region expression for the $F_r/q$ can be presented as follows:

\begin{equation}\label{eq:17}
\begin{split}
&\frac{{{F_r}(r < {r_0},\varphi ,\xi ,{r_0},{\varphi _0},{t_0})}}{q} = 2qk_p^2\frac{{{I'_m}({k_p}r)}}{{{I_m}({k_p}a)}}\times\\
&{\Delta _m}({k_p}a,{k_p}{r_0})\sin {\omega _p}(\xi  - {t_0})\theta (\xi  - {t_0}){e^{ - im{\varphi _0}}} + \\
&\sum\limits_{s = 1} {\frac{{4q{\kappa _{ps}}v{D_2}({\omega _s})}}{{a\omega _s^2D'({\omega _s})}}} \frac{{{I_m}({\kappa _{ps}}{r_0}){I'_m}({\kappa _{ps}}r)}}{{I_m^2({\kappa _{ps}}a)}} \times \\
&\sin {\omega _s}(\xi  - {t_0})\theta (\xi  - {t_0}){e^{ - im{\varphi _0}}},
\end{split}
\end{equation}

\begin{equation}\label{eq:18}
\begin{split}
&\frac{{{F_r}({r_0} < r \le a,\varphi ,\xi ,{r_0},{\varphi _0},{t_0})}}{q} = 2qk_p^2\frac{{{I_m}({k_p}{r_0})}}{{{I_m}({k_p}a)}} \times \\
&{\Delta '_m}({k_p}a,{k_p}r)\sin {\omega _p}(\xi  - {t_0})\theta (\xi  - {t_0}){e^{ - im{\varphi _0}}} + \\
&\sum\limits_{s = 1} {\frac{{4q{\kappa _{ps}}v{D_2}({\omega _s})}}{{a\omega _s^2D'({\omega _s})}}} \frac{{{I_m}({\kappa _{ps}}{r_0})I'_m({\kappa _{ps}}r)}}{{I_m^2({\kappa _{ps}}a)}} \times \\
&\sin {\omega _s}(\xi  - {t_0})\theta (\xi  - {t_0}){e^{ - im{\varphi _0}}}
\end{split}
\end{equation}

In order to obtain the wakefield excited by a drive bunch of finite size, it is necessary to integrate the expressions for the point--like particle over the arrival time $t_0$ and by the transverse coordinates $r_0$, and $\varphi_0$. It can be supposed that: (I) the drive bunch has a square profile of the charge density in both longitudinal and transverse directions (uniform distribution), (II) the bunch offset $R_{off}$ is greater than the drive bunch radius $R_b$. The resulting expressions for the axial and radial forces acting on the particles, which are located near the plasma--dielectric interface ($r=R_{off} + R_b$) and accelerating structure axis ($r=R_{off} - R_b$) can be presented as follows:

\begin{equation}\label{eq:19}
\begin{split}
&\frac{{{F_z}(r = R_{off} + R_b,\varphi ,\xi )}}{q} =  - \frac{{4{Q_b}}}{{{R_b}{L_b}}}\sum\limits_{m =  - \infty }^{ + \infty } {{e^{im\varphi }}}  \times \\
&\bigg( {\frac{{{I_m}({k_p}{R_{off}}){I_1}({k_p}{R_b})}}{{{I_m}({k_p}a)}}} {\Delta _m}({k_p}a,{k_p}({R_{off}} + {R_b}))\Psi _\parallel ^{(p)}(\xi ) + \\
&\sum\limits_{s = 1} {\frac{{2v{D_2}({\omega _s})}}{{a{\kappa _{ps}}\omega _s^2D'({\omega _s})}}\frac{{{I_m}({\kappa _{ps}}{R_{off}}){I_1}({\kappa _{ps}}{R_b})}}{{I_m^2({\kappa _{ps}}a)}}}  \times \\
& {{I_m}({\kappa _{ps}}({R_{off}} + {R_b}))\Psi _\parallel ^{(s)}(\xi )} \bigg)
\end{split}
\end{equation}

\begin{equation}\label{eq:20}
\begin{split}
&\frac{{{F_r}(r = {R_{off}} + {R_b},\varphi ,\xi )}}{q} = \frac{{4{Q_b}}}{{{R_b}{L_b}}}\sum\limits_{m =  - \infty }^{ + \infty } {{e^{im\varphi }} \times } \\
&\bigg(\frac{{{I_m}({k_p}{R_{off}}){I_1}({k_p}{R_b})}}{{{I_m}({k_p}a)}}{\Delta '_m}({k_p}a,{k_p}({R_{off}} + {R_b}))\Psi _ \bot ^{(p)}(\xi ) + \\
&\sum\limits_{s = 1} {\frac{{2{v^2}{D_2}({\omega _s})}}{{a\omega _s^3D'({\omega _s})}}\frac{{{I_m}({\kappa _{ps}}{R_{off}}){I_1}({\kappa _{ps}}{R_b})}}{{I_m^2({\kappa _{ps}}a)}}}  \times \\
&{I'_m}({\kappa _{ps}}({R_{off}} + {R_b}))\Psi _ \bot ^{(s)}(\xi )\bigg),
\end{split}
\end{equation}

\begin{equation}\label{eq:21}
\begin{split}
&\frac{{{F_z}(r = {R_{off}} - {R_b},\varphi ,\xi )}}{q} =  - \frac{{4{Q_b}}}{{{R_b}{L_b}}}\sum\limits_{m =  - \infty }^{ + \infty } {{e^{im\varphi }}}  \times \\
&\bigg(\frac{{{I_m}({k_p}({R_{off}} - {R_b})){I_1}({k_p}{R_b})}}{{{I_m}({k_p}a)}}{\Delta _m}({k_p}a,{k_p}{R_{off}})\Psi _\parallel ^{(p)}(\xi ) + \\
&\sum\limits_{s = 1} {\frac{{2v{D_2}({\omega _s})}}{{a{\kappa _{ps}}\omega _s^2D'({\omega _s})}}\frac{{{I_m}({\kappa _{ps}}{R_{off}}){I_1}({\kappa _{ps}}{R_b})}}{{I_m^2({\kappa _{ps}}a)}}}  \times \\
&{I_m}({\kappa _{ps}}({R_{off}} - {R_b}))\Psi _\parallel ^{(s)}(\xi )\bigg),
\end{split}
\end{equation}

\begin{equation}\label{eq:22}
\begin{split}
&\frac{{{F_r}(r = {R_{off}} - {R_b},\varphi ,\xi )}}{q} = \frac{{4{Q_b}}}{{{R_b}{L_b}}}\sum\limits_{m =  - \infty }^{ + \infty } {{e^{im\varphi }}}  \times \\
&\bigg(\frac{{{I'_m}({k_p}({R_{off}} - {R_b})){I_1}({k_p}{R_b})}}{{{I_m}({k_p}a)}}{\Delta _m}({k_p}a,{k_p}{R_{off}})\Psi _ \bot ^{(p)}(\xi ) + \\
&\sum\limits_{s = 1} {\frac{{2{v^2}{D_2}({\omega _s})}}{{a\omega _s^3D'({\omega _s})}}\frac{{{I_m}({\kappa _{ps}}{R_{off}}){I_1}({\kappa _{ps}}{R_b})}}{{I_m^2({\kappa _{ps}}a)}}}  \times \\
&{I'_m}({\kappa _{ps}}({R_{off}} - {R_b}))\Psi _ \bot ^{(s)}(\xi )\bigg),
\end{split}
\end{equation}
where $Q_b$ is the bunch charge, and $L_b$ is the bunch length. It is important to note that for the last integration, the method of waveguide structure excitation by the off--axis beam, well known in the theory of gyrotrons, was used. The principle of this integration method is the transition from the laboratory frame of reference to the frame of reference with the origin located at the center of the bunch (a detailed description of the procedure can be found in Appendix). A detailed description of this method can be found, for example, in paper~\cite{Chu1988}.

The axial structure of the wakefield is described by the functions $\Psi_{\parallel}^{p,s}(\xi)$ and $\Psi_{\perp}^{p,s}(\xi)$:
\begin{equation}\label{eq:23}
\begin{split}
\Psi_{\parallel}^{p,s}(\xi) = &{\theta (\xi)\sin {\omega _{p,s}}\xi}-\\
&{\theta (\xi - L_b/v)\sin {\omega _{p,s}}(\xi - L_b/v)},\\
\Psi_{\perp}^{p,s}(\xi) = &\theta (\xi)(1 - \cos {\omega _{p,s}}\xi)-\\
&\theta (\xi - L_b/v)(1 - \cos {\omega _{p,s}}(\xi - L_b/v)).
\end{split}
\end{equation}
The analytical expressions~(\ref{eq:19})--(\ref{eq:22}) allow carrying out a numerical analysis of the amplitude distribution of the axial and radial forces acting on the particles that are on the envelope of the drive bunch, which is injected with offset.

\section{NUMERICAL ANALYSIS RESULTS}
Based on the derived wakefield excitation theory, a numerical analysis was performed for the waveguide parameters in the terahertz frequency range; for the drive bunch we have used parameters of electron bunches accessible at SLAC. These parameters are presented in Table~\ref{Tabl_1}.
\begin{table}[!h]
   \centering
   \caption{Parameters of the plasma--dielectric waveguide and the drive electron bunch}
   \begin{tabular}{lcc}
   \hline
   \hline
       \textbf{Parameter} & \textbf{Value}                             \\
   \hline
   \hline
          Inner dielectric radius        & $0.5\,mm$                   \\
          Outer dielectric radius        & $0.6\,mm$                   \\
          Permittivity of the dielectric & $3.75 (quartz)$              \\
          Plasma density                 & $4.41\cdot 10^{14}\,cm^{-3}$\\
          Energy of bunch                & $5\,GeV$                    \\
          Charge of bunch                & $3\,nC$                     \\
          Length of bunch                & $0.5\,mm$                   \\
          Radius of bunch                & $0.23\,mm$                  \\
          Bunch offset                   & $0.24\,mm$                  \\
   \hline
   \hline
   \end{tabular}
   \label{Tabl_1}
\end{table}
First of all, the numerical methods were used to solve dispersion equations for resonant $TM$-modes (which are working modes for the accelerator structures) and hybrid electromagnetic $HEM$-modes (which are parasitic, because their excitation causes the BBU instability). For the parameters of the waveguide under study and the electron 5 GeV drive bunch, the resonant frequencies of the first six eigen $TM$-modes are equal to 0.357 THz, 1.160 THz, 1.898 THz, 2.774 THz, 3.663 THz and 4.558 THz, respectively. The resonant frequencies of the first six eigen $HEM$-modes are 0.025 THz, 0.356 THz, 0.532 THz, 1.066 THz, 1.387 THz and 1.903 THz, respectively.

\begin{figure}[!th]
  \centering
  \includegraphics[width=0.5\textwidth]{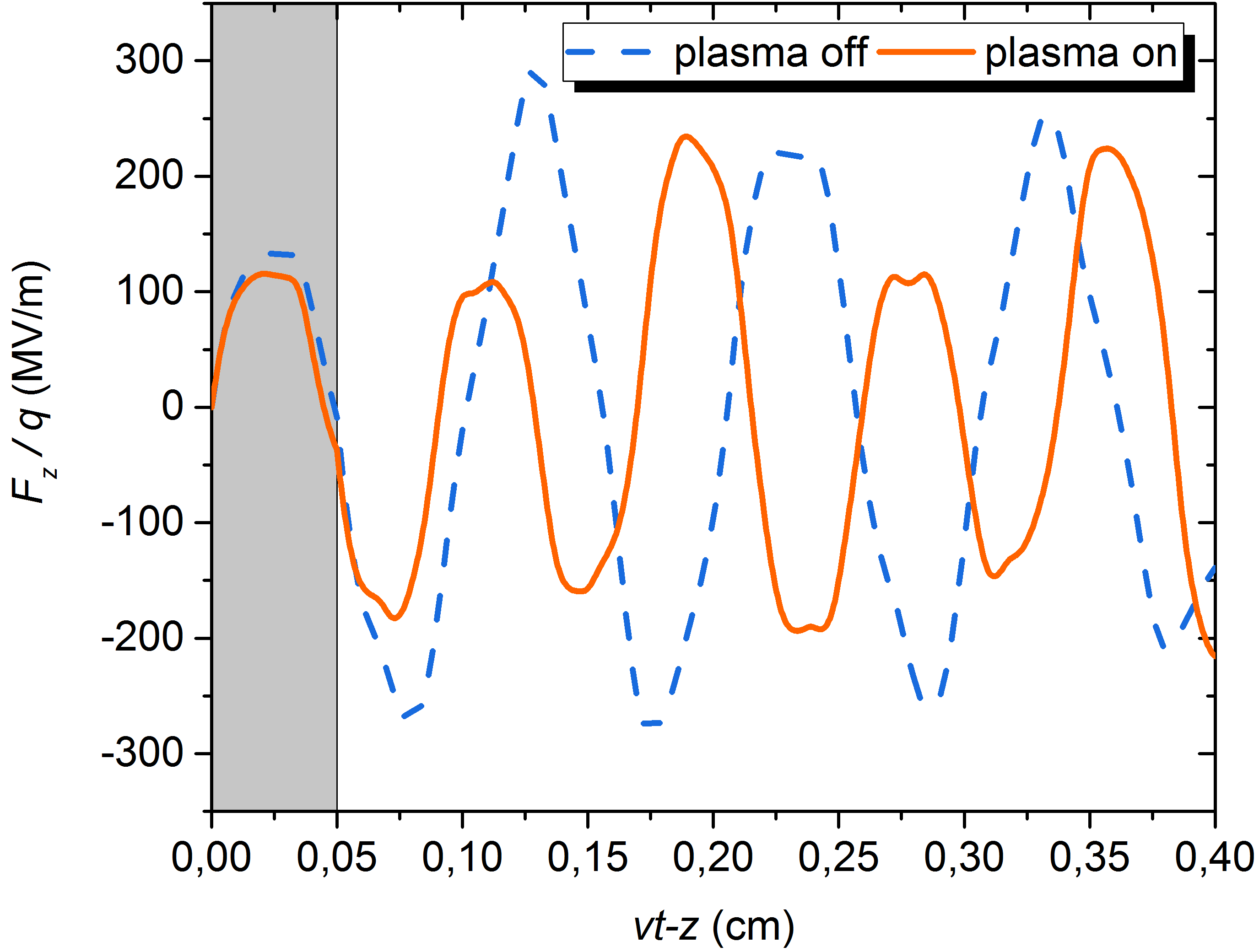}
  \includegraphics[width=0.5\textwidth]{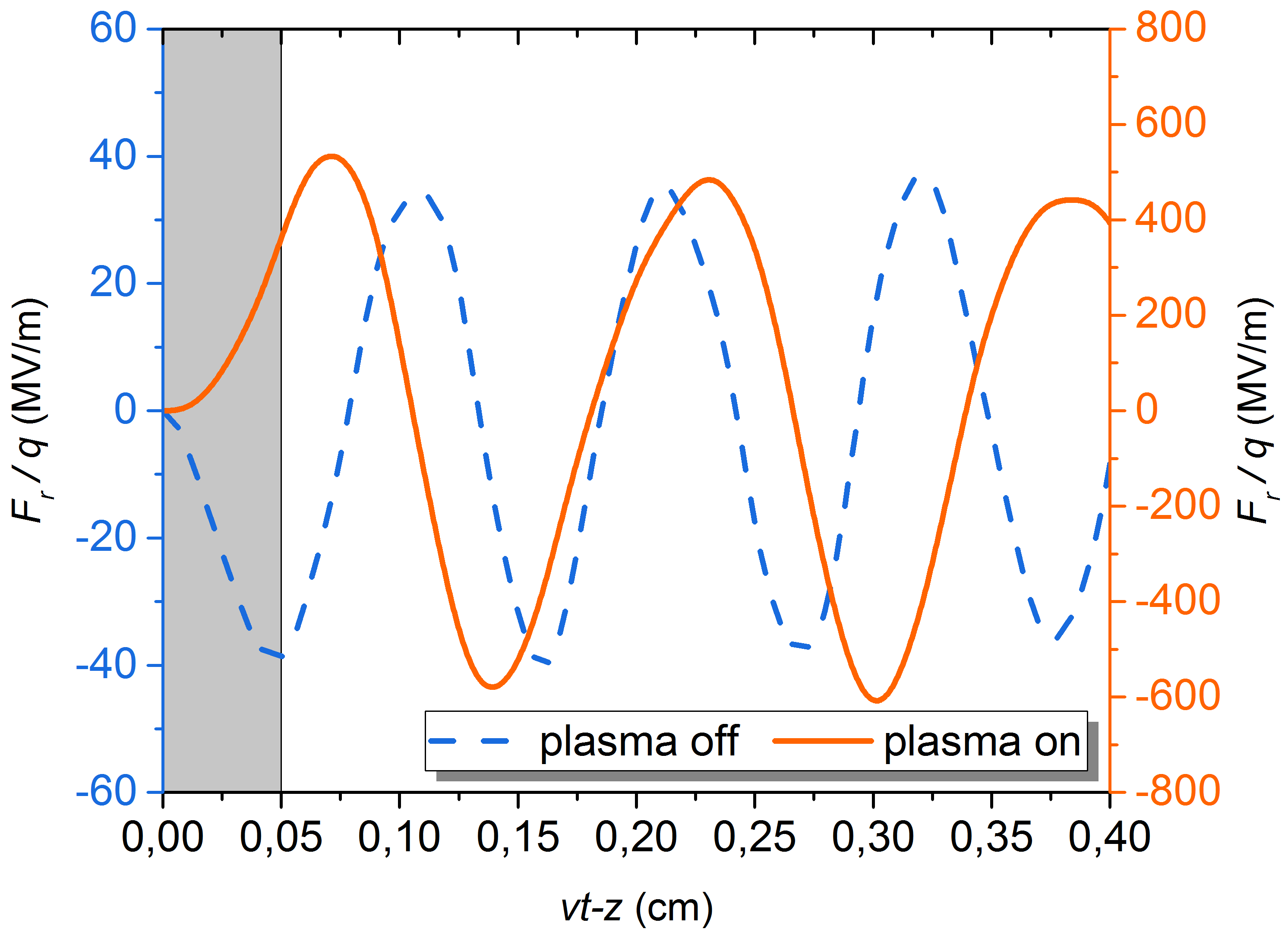}
  \caption{Longitudinal profiles of the axial and radial forces at a distance of 0.47 mm from the waveguide axis for the plasma--filled channel (solid line), and in the absence of plasma in the channel (dashed line). The drive bunch (grey rectangle) head position corresponds to $vt-z=0$.}\label{Fig:02}
\end{figure}

\begin{figure}[]
  \centering
  \includegraphics[width=0.5\textwidth]{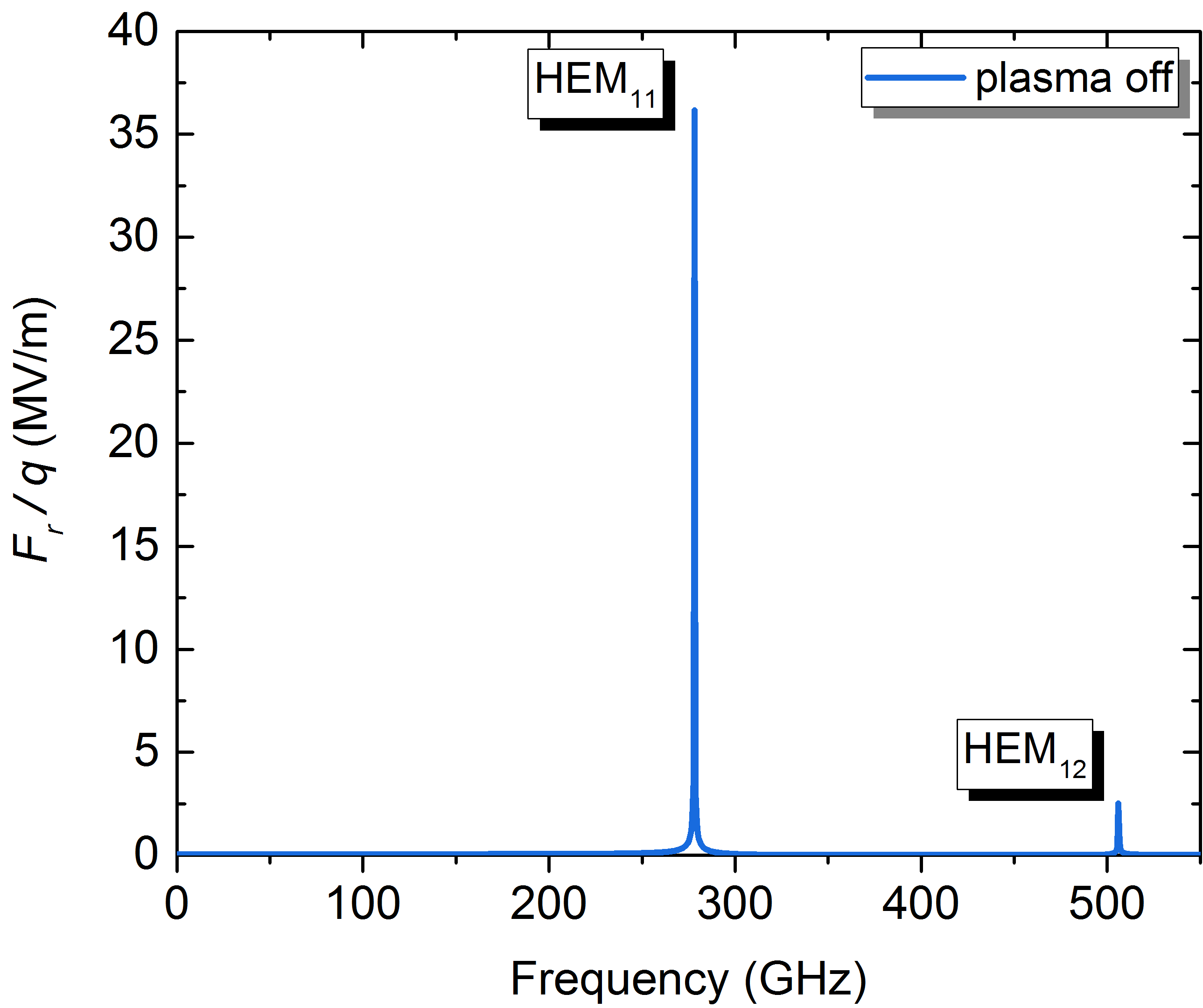}
  \includegraphics[width=0.5\textwidth]{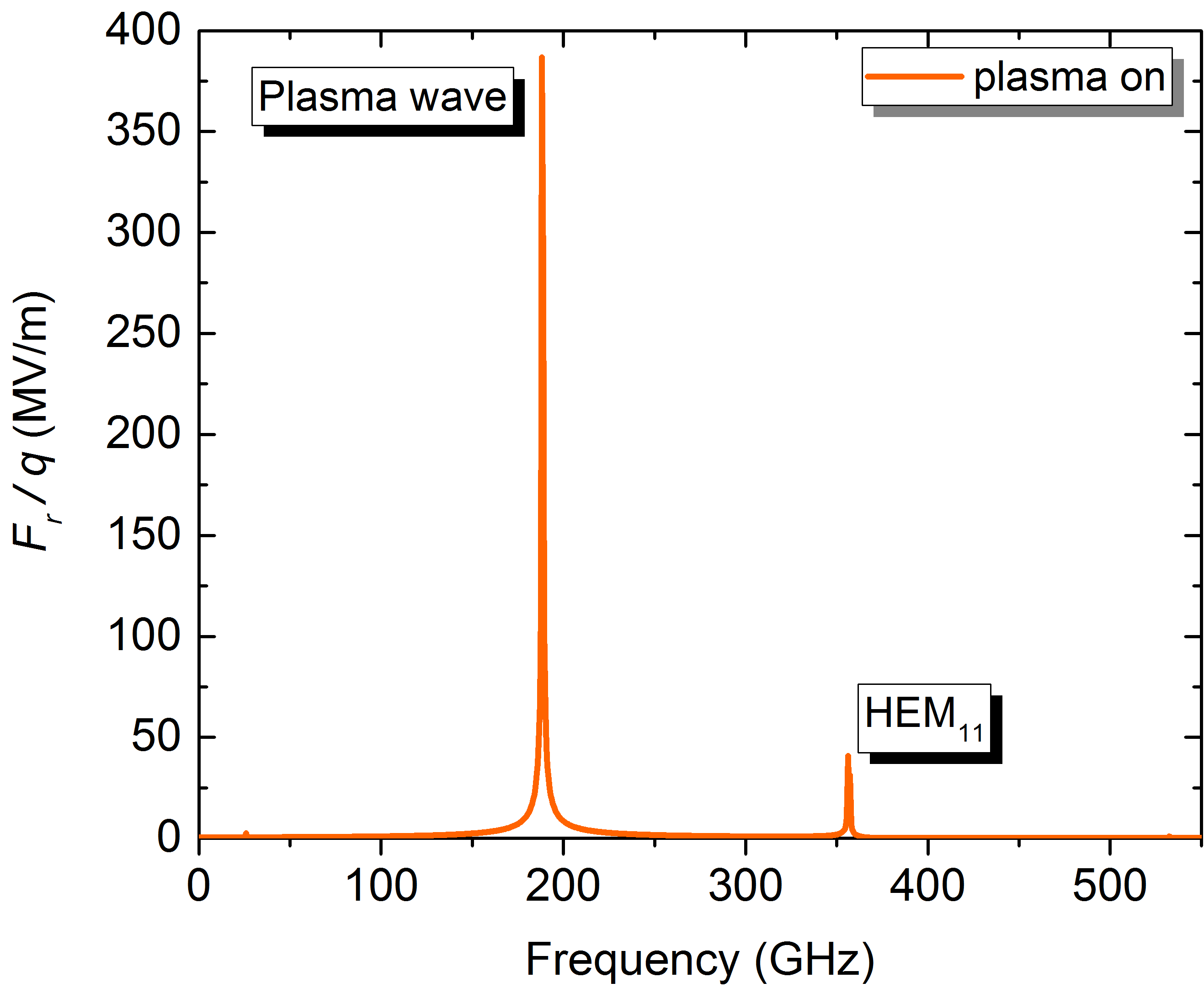}
  \caption{Radial force spectrum (at $r = 0.47 mm$ ) excited by the off--axis drive electron bunch for the plasma off (top) and plasma on (bottom) cases.}\label{Fig:03}
\end{figure}

\begin{figure}[!h]
  \centering
  \includegraphics[width=0.5\textwidth]{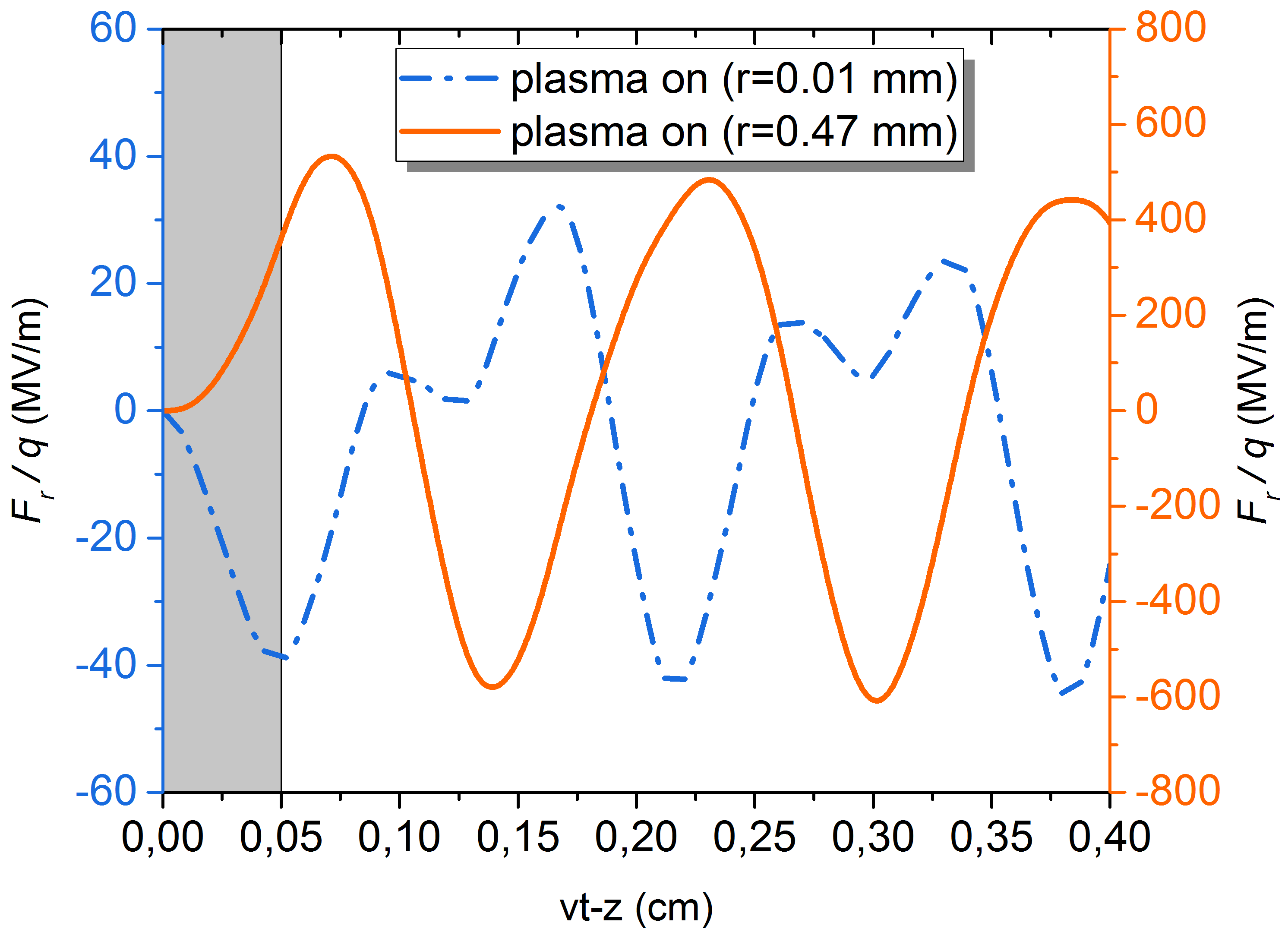}\hfill
  \caption{Axial profiles of the radial force at distances of 0.47 mm (solid line), and 0.01 mm (dashed line) from the waveguide axis. The drive bunch (grey rectangle) head position corresponds to $vt-z=0$.}\label{Fig:04}
\end{figure}

The presence of plasma in the channel for charged particles leads to a change in both the amplitudes of the bunch--excited wakefield components and their spatial distribution. The axial profiles of the axial and radial forces were compared for the case of filling the channel for charged particles with plasma and the case of no plasma in the channel. The results of this comparison are presented in \Figref{Fig:02} as the longitudinal distributions of axial and radial forces that acting on its particles.
Axial distributions in \Figref{Fig:02} are shown along a line which is parallel to the structure axis and located at a distances of 0.47 mm from it, corresponding to the envelope of the drive bunch closer to the dielectric. Figure~\ref{Fig:02} illustrates, that the axial force amplitude in the plasma--dielectric structure is less than in the corresponding dielectric--loaded structure behind the drive bunch. Filling the channel with plasma also leads to decrease in the spatial period of the axial force. An analysis showed, that the above changes are associated with the excitation of the longitudinal electric field of the plasma wave, as well as with the change in the frequencies of the plasma--dielectric waveguide eigenwaves. As we are mainly interested in the BBU of the drive bunch, we start with analysis of the transverse wakefield amplitudes. As can be seen from \Figref{Fig:02}, the radial force profile shows almost a harmonic dependence on the longitudinal coordinate, with a period approximately equal to the plasma wavelength with a low--frequency modulation. It can be seen that filling with plasma causes a qualitative change in the longitudinal profile of the radial force. This change is especially noticeable in the the bunch region, where the radial force has changed its sign. Namely, in the absence of plasma in the channel for charged particles, the radial force is defocusing, but in the presence of plasma, this force becomes focusing. In turn, this should be accompanied by the change in the frequency composition of the excited wakefield. Therefore, the spectral analysis of the off--axis bunch--excited wakefield was carried out for the cases of plasma off and plasma on. The results of the spectral analysis are presented in \Figref{Fig:03}.
The analysis shows that in the absence of plasma the off--axis drive bunch leads to significant excitation of $HEM$--modes with the azimuthal index $m=1$ (dipole modes). The analysis has shown that these eigenmodes make a dominant contribution to the radial force, under the action of which the drive bunch particles will gradually move to the dielectric. Comparison of spectra demonstrates that the main reason of difference between the amplitudes and axial profiles of the radial forces for cases of presence and absence of the plasma lies in the plasma wave excitation by the drive bunch. The radial electric field of this wave acts on the bunch particles as a focusing system. Thus the plasma wave makes a predominant contribution to the radial force. The low--frequency modulation of the radial force amplitude is due to excitation of the eigenwaves of the plasma--dielectric accelerating structure. This change in the excited wakefield spectrum leads, in turn, to a significant (qualitative and quantitative) change in the transverse dynamics of the drive bunch. The classical scenario of the BBU instability is as follows. As a result of the electron bunch injection with offset the bunch tail is deflected during passing down the accelerating structure. A physical reason of this deflection is a significant transverse wakefield component excitation. The presence of plasma in the channel leads to the change of the aforementioned scenario. Figure~\ref{Fig:04} presents the distribution of radial forces, which are acting on its peripheral particles (corresponding, respectively to the drive bunch envelope closer to the dielectric, and the drive bunch envelope closer to the waveguide axis).
Figure~\ref{Fig:04} indicates that the radial dynamics of the drive bunch particles located on the bunch opposite transverse peripheries will differ both qualitatively and quantitatively. The bunch particles which are closer to the plasma--dielectric interface will focus in the axial direction during their propagation along the accelerating structure. The focusing force increases along the bunch from its head to tail. On the contrary, the bunch particles which are closer to the waveguide axis will defocus during propagation. The defocusing force also increases along the bunch from its its head to tail. As a result, in the process of wakefield excitation the off--axis drive bunch will not deposit on the dielectric surface. And instead of a gradual increase of transverse displacement, there will be radial compression of the tail end of the bunch.

\begin{figure}[h]
  \centering
  \includegraphics[width=0.5\textwidth]{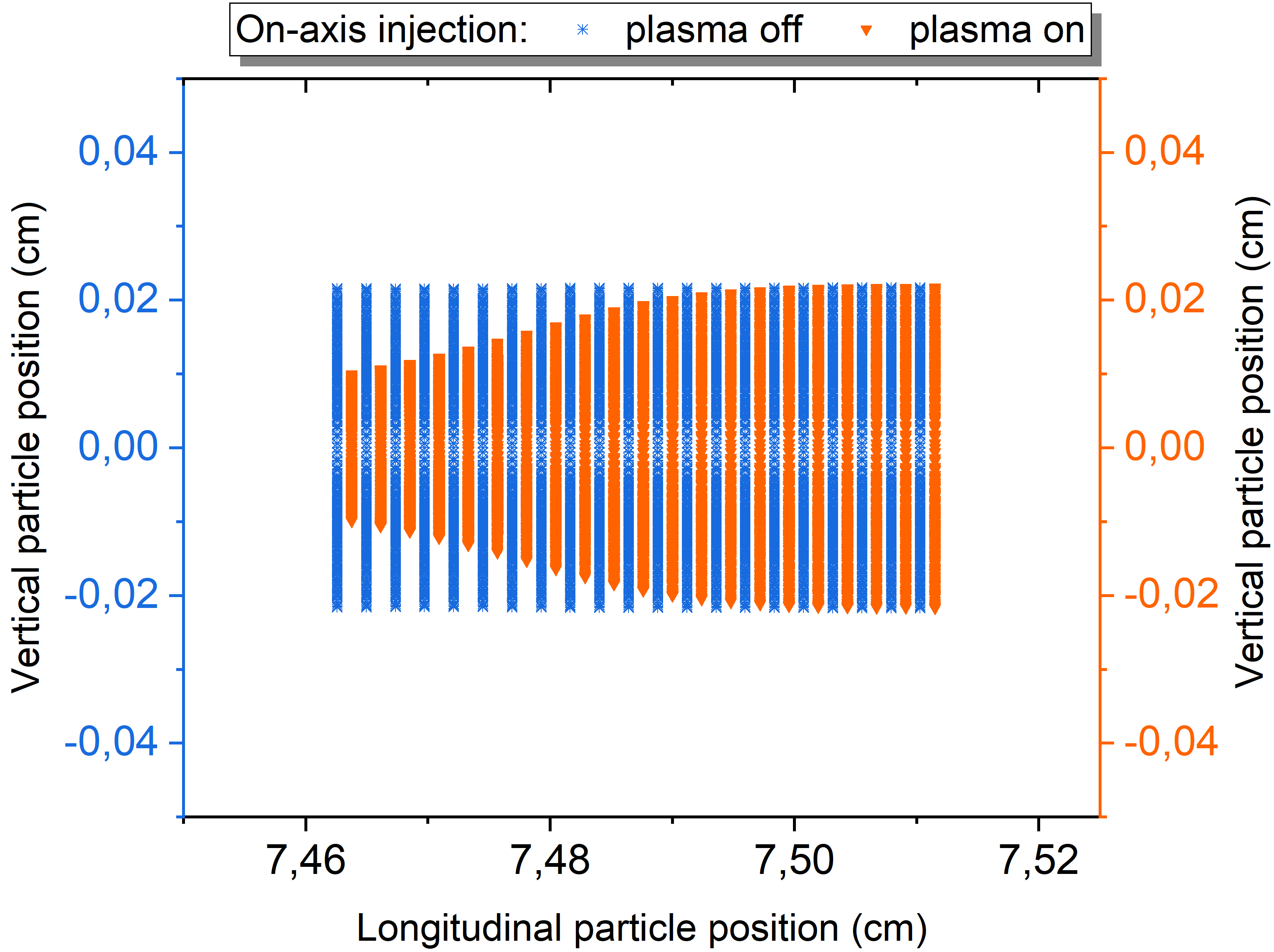}\hfill
  \caption{The on--axis injected drive bunch particle distribution on the vertical--longitudinal plane at the accelerating structure exit for the plasma on--and plasma off--cases. The vertical axis limits correspond to the full width of the channel for the charged particles, the horizontal axis limits correspond to the frame around the bunch.}\label{Fig:05}
\end{figure}

\begin{figure}[h]
  \centering
  \includegraphics[width=0.5\textwidth]{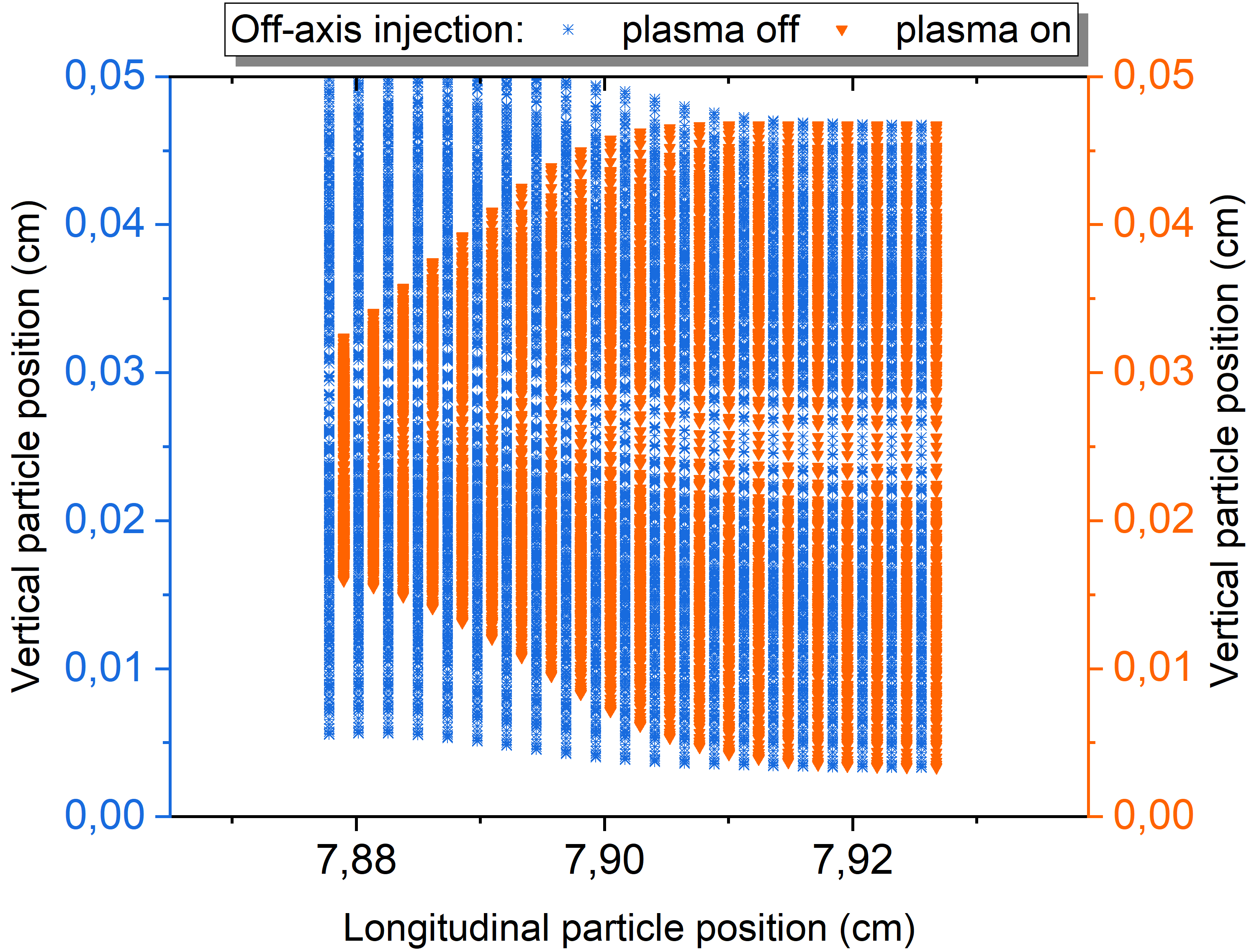}\hfill
  \caption{The off--axis injected drive bunch particle distribution on the vertical--longitudinal plane at the accelerating structure exit for the plasma on and plasma off cases. The vertical axis limits correspond to the half width of channel for the charged particles, horizontal axis limits correspond to the frame around the bunch.}\label{Fig:06}
\end{figure}

 By using the profiles of the transverse force that acts on the drive bunch particles, the developed theory predicts the transverse dynamics of the drive bunch. To verify our wakefield amplitude calculations, which are based on analytical expressions, the particle-in-cell simulations of the drive electron bunch propagation along the structure under study were carried out, too. The main goal of the simulations was to verify the derived theory predictions. For simulations, we have used the same parameters of the waveguide and drive bunch as those for the analytical calculations (see Table~\ref{Tabl_1}). Both the plasma off and plasma on cases for the off-axis injection were analyzed. We started the simulations with the case of on--axis injection.  Figure~\ref{Fig:05} demonstrates the positions of the particles for the time moment when the drive bunch reached the longitudinal position $z=7.51 cm$. The positions of the particles are given in the vertical--longitudinal plane. It is evident from the figure that the presence of plasma results in focusing of particles located in the tail of the drive bunch.
 It should be noted, that the simulated  "model" monoenergetic drive bunch has zero divergence at the structure input. For the nonzero divergence drive bunch, the focusing features of the plasma--dielectric structure can partially compensate the divergence increase in the process of bunch propagation, and will be investigated further. Figure~\ref{Fig:06} shows the bunch particles positions on the vertical--longitudinal plane at the time when the bunch started to deposit on the dielectric in the plasma off case.
As it is shown, there is a significant qualitative difference in the transverse bunch dynamics in the plasma off and plasma on cases. As a result of the initial offset, the transverse deflection of the tail particles is greater than that of the head ones in the absence of plasma in the channel for the charged particles. The drive bunch is deflected as a whole while it propagates through the dielectric--loaded structure, and the BBU instability does exist in this case, causing undesirable bunch displacement, to occur  up to the total loss of the bunch. The presence of plasma in the channel for the charged particles leads to the change in the transverse particle dynamics. The particles in the tail of the off-axis drive bunch are focused (instead of deflection) due to plasma wave excitation. As a result, the bunch offset does not tend to increase with time and there is no gradual increase of charge losses does not occur. The results of numerical simulations show a good qualitative agreement with the corresponding analytical theory predictions in terms of the drive bunch envelope particles behavior. More detailed numerical simulations will be performed during further studies. Thus, it may be concluded that the presence of plasma in the channel for charged particles leads to suppression of the transverse instability of the drive bunch. In turn, from a practical standpoint, this will enable us: (I) to prevent the charge losses of the drive bunch, which excites the wakefield, (II) to reduce the requirements for the accuracy of the drive bunch injection into the accelerating structure, this being particularly important for terahertz structures having submillimeter transverse dimensions, (III)to  increase the acceleration length of the witness bunch and, consequently, to increase its output energy.
In conclusion, it may be said, that the intrinsic focusing features of such plasma--dielectric accelerating structure under discussion allow to improve the drive bunch stability during wakefield excitation.

\section{CONCLUSIONS}
An analytical theory of wakefield excitation by the off-axis particle bunch in the round plasma--dielectric waveguide has been formulated. A comparison between the longitudinal profiles of radial forces, excited by the off--axis drive electron bunch, has been carried out for the cases of the absence and presence of plasma in the channel for charged particles. It has been demonstrated that in contrast to the dielectric--loaded structure without plasma filling, the presence of the initial bunch offset in the plasma--dielectric accelerating structure does not cause to the beam breakup instability.

\begin{acknowledgments}
The study is supported by the National Research Foundation of Ukraine under the program “Leading and Young Scientists Research Support” (project \# 2020.02/0299).
\end{acknowledgments}

\section*{APPENDIX: INTEGRATION OVER THE PARTICLES}
Consider following integral over the arrival time $t_0$ and transverse coordinates $r_0$, and $\varphi_0$
\begin{equation}\label{eq:24}
\begin{split}
\int\int\int f(r,\varphi,\xi,r_{0},\varphi_{0},t_{0})dt_{0}dr_{0}d\varphi_{0},
\end{split}
\end{equation}
here $f(r,\varphi,\xi,r_{0},\varphi_{0},t_{0})$ is a function, which represents any of the wakefield components, excited by the point--like particle (Green's function). To calculate this integral it is more convenient to transit from the laboratory frame of reference to the frame of reference with the origin, located at the center of the drive bunch.
\begin{figure}[h]
  \centering
  \includegraphics[width=0.5\textwidth]{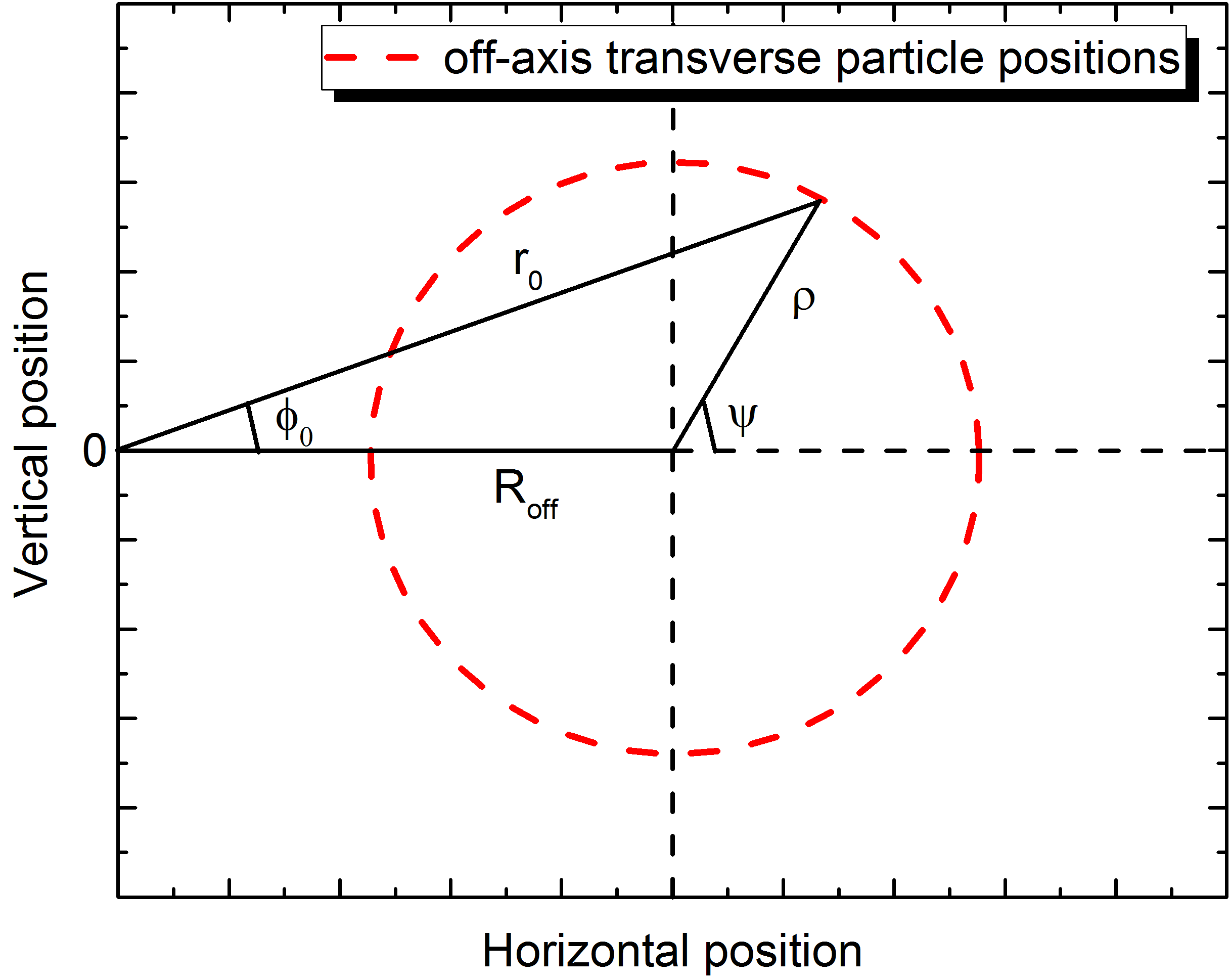}\hfill
  \caption{The off--axis injected drive bunch particle positions on horizontal--vertical plane. Particle transverse coordinates $r_0$, and $\varphi_0$ in the laboratory frame of reference, and transverse coordinates $\rho$ and $\psi$ in the bunch frame of reference are shown schematically.}\label{Fig:07}
\end{figure}
Particle transverse coordinates $r_0$, and $\varphi_0$ in the laboratory frame of reference are expressed in terms of new corresponding transverse coordinates $\rho$ and $\psi$ in the bunch frame of reference by the expressions
\begin{equation}\label{eq:25}
\begin{split}
&r_{0}(\rho,\psi)=\sqrt{R_{off}^2 + 2R_{off}\rho cos\psi + \rho^2},\\
&tg\varphi_{0}(\rho,\psi)=\frac{\rho sin\psi}{R_{off} + \rho cos\psi}
\end{split}
\end{equation}
For the sake of simplicity we present horizontal drive bunch offset only.

Using new variables $\rho$ and $\psi$, one can rewrite the integral~(\ref{eq:24}) as
\begin{equation}\label{eq:26}
\begin{split}
\int\int\int &f(r,\varphi,\xi,r_{0},\varphi_{0},t_{0})dt_{0}dr_{0}d\varphi_{0}=\\
\int\int\int &f(r,\varphi,\xi,r_{0}(\rho,\psi),\varphi_{0}(\rho,\psi),t_{0})\times\\
&J(r_{0},\varphi_{0};\rho,\psi)dt_{0}d\rho d\psi,
\end{split}
\end{equation}
here $J(r_{0},\varphi_{0};\rho,\psi)$ is the Jacobian determinant.

The Jacobian determinant defines as
\begin{equation}\label{eq:27}
J(r_{0},\varphi_{0};\rho,\psi)=\bigg(\frac{\partial r_{0}}{\partial \rho}\bigg)\bigg(\frac{\partial \varphi_{0}}{\partial \psi}\bigg) - \bigg(\frac{\partial r_{0}}{\partial \psi}\bigg)\bigg(\frac{\partial \varphi_{0}}{\partial \rho}\bigg),
\end{equation}
and substituting expressions~(\ref{eq:25}) into~(\ref{eq:27}), one obtains
\begin{equation}\label{eq:28}
J(r_{0},\varphi_{0};\rho,\psi)=\frac{\rho}{r_{0}}.
\end{equation}

Function $f(r,\varphi,\xi,r_{0},\varphi_{0},t_{0})$ contains the combinations $I_{m}(k_{p}r_{0})e^{-im\varphi_{0}}$ and $K_{m}(k_{p}r_{0})e^{-im\varphi_{0}}$ (see the expressions ~(\ref{eq:9})--~(\ref{eq:16})). Using the Graph theorem for the cylindrical functions~\cite{Bateman} it is possible to rewrite these combinations in terms of new variables $\rho$ and $\psi$, as follows:
\begin{equation}\label{eq:29}
\begin{split}
I_{m}(k_{p}r_{0})e^{-im\varphi_{0}}=\sum\limits_{\nu=-\infty}^{+\infty}&(-1)^\nu I_{m+\nu}(k_{p}R_{off})I_{\nu}(k_{p}\rho)\times \\
&e^{-i\nu(\pi-\psi)}, \\
K_{m}(k_{p}r_{0})e^{-im\varphi_{0}}=\sum\limits_{\nu=-\infty}^{+\infty}&K_{m+\nu}(k_{p}R_{off})I_{\nu}(k_{p}\rho)\times \\
&e^{-i\nu(\pi-\psi)}
\end{split}
\end{equation}
Since we supposed, that the drive bunch has uniform distribution of the charge density in both longitudinal and transverse directions, the charge of the point--like particle is
\begin{equation}\label{eq:30}
q=Q_{b}\frac{r_{0}dr_{0}d\varphi_{0}vdt_{0}}{\pi R_b^2L_b}
\end{equation}
Now, one can accomplish the integration over the arrival time and transverse coordinates of the particles to get the wakefield, excited by the drive bunch of finite size.

\bibliography{Bibliography}	
\end{document}